\documentclass{article}
\usepackage{spconf,amsmath,graphicx}

\usepackage{algorithm}
\usepackage{algpseudocode}

\usepackage{multirow}
\usepackage[table,xcdraw]{xcolor}

\title{Improving Federated Learning Communication Efficiency with Global Momentum Fusion for Gradient Compression Schemes}
\name{
    Chun-Chih Kuo$^{1,3}$,
    Ted Tsei Kuo$^{1,*}$ \thanks{$*$ corresponding author},
    Chia-Yu Lin$^{2}$}

\address{$^{1}$Institute of Computational Intelligence, National Yang Ming Chiao Tung University\\
$^{2}$Department of Computer Science and Information Engineering National Central University\\
$^{3}$AI Lab, Trend Micro\\
}
\begin{document}
%\ninept
%
\maketitle
\begin{abstract}
Communication costs within Federated learning hinder the system scalability for reaching more data from more clients. The proposed FL adopts a hub-and-spoke network topology.  All clients communicate through the central server. Hence, reducing communication overheads via techniques such as data compression has been proposed to mitigate this issue. Another challenge of federated learning is unbalanced data distribution, data on each client are not independent and identically distributed (non-IID) in a typical federated learning setting. In this paper, we proposed a new compression compensation scheme called Global Momentum Fusion (GMF) which reduces communication overheads between FL clients and the server and maintains comparable model accuracy in the presence of non-IID data.

\end{abstract}
\section{Introduction}
\label{sec:intro}

Federated learning, proposed by McMahan et al.\cite{mcmahan2017communication}, provides a new paradigm to enhance the privacy-promised learning framework without moving data outside the devices. Federated learning allows clients to train models with personal data on their devices and only share intermediate gradients with the central server. The server aggregates these gradients (typically by average) and transmits aggregated gradient back to clients. Finally, clients update their models as the new base model for the next round.

While clients communicate through the central server, the computation and communication of the server have become system overheads. Reducing communication overheads via techniques such as gradient compression has been proposed to mitigate this issue. Most compression techniques are considered lossy \cite{xu2020compressed}. Thus, some compensation techniques are usually required to adopt memory mechanism. For example, Lin et al. proposed Deep Gradient Compression (DGC) \cite{lin2018deep} to achieve a high compression ratio via well compensation design. However, DGC did not consider the non-IID problem discussed below. Another challenge of Federated Learning is unbalanced data distribution. Unlike centralized distributed learning, which separates data with balance distribution, data on each client are not independent and identically distributed (non-IID) in a typical Federated Learning setting. Several approaches have been proposed using global momentum methods to mitigate this problem, such as Sparse Gradient Compression \cite{aji2017sparse} and Global Momentum Compression \cite{zhao2019global}. However, these solutions further increased communication overheads and decreased accuracy.  Although these existing methods showed fine in performance in different consideration, thus our main motivation is to design a scheme which cloud address both communication overheads and non-IID data. 

This paper proposes two main contributions. First, we analyze two main communication overheads cost with the existing method. We theoretically show that some methods using momentum could lead to additional communication overheads. Second, we proposed a new compression compensation scheme called Global Momentum Fusion (GMF) which reduces communication overheads between Federated learning clients and the server and maintains comparable model accuracy in the presence of non-IID data.

% We organize this thesis as follows. We first introduce related work in Chapter \ref{CHA:2}. Next, in Chapter \ref{CHA:3}, we formulated the problem with references to the approaches of some prior work and proposed a GMF compression technique to address the focused challenges. In Chapter \ref{CHA:4}, we provide reproducible experimental results to compare the performance of our proposed method, GMF, against others’ approaches’ results. Finally, a conclusion is given in Chapter \ref{CHA:5}.``

% These guidelines include complete descriptions of the fonts, spacing, and
% related information for producing your proceedings manuscripts. Please follow
% them and if you have any questions, direct them to Conference Management
% Services, Inc.: Phone +1-979-846-6800 or email
% to \\\texttt{papers@2021.ieeeicassp.org}.

% \section{Problem Formulation}
% \label{sec:format}

% All printed material, including text, illustrations, and charts, must be kept
% within a print area of 7 inches (178 mm) wide by 9 inches (229 mm) high. Do
% not write or print anything outside the print area. The top margin must be 1
% inch (25 mm), except for the title page, and the left margin must be 0.75 inch
% (19 mm).  All {\it text} must be in a two-column format. Columns are to be 3.39
% inches (86 mm) wide, with a 0.24 inch (6 mm) space between them. Text must be
% fully justified.

\section{Problem formulation}
Based on our study, we illustrate two major problems of existing approaches.

\subsection{Server-Side Global Momentum Leads to Extra Communication Overheads}\label{PF:Server-Side}
Communication overheads comprise two main traffic costs. ($i$) clients upload gradients to the server to aggregate, and ($ii$) the server transmits the aggregated gradient to the clients. The size of the gradient upload from clients has a fixed size. However, the size of the aggregated gradient could be varied since the gradients collected from clients are inconsistent.

\begin{figure}[htb]
\vspace{-0em}    
\centering  
\includegraphics[width=8cm]{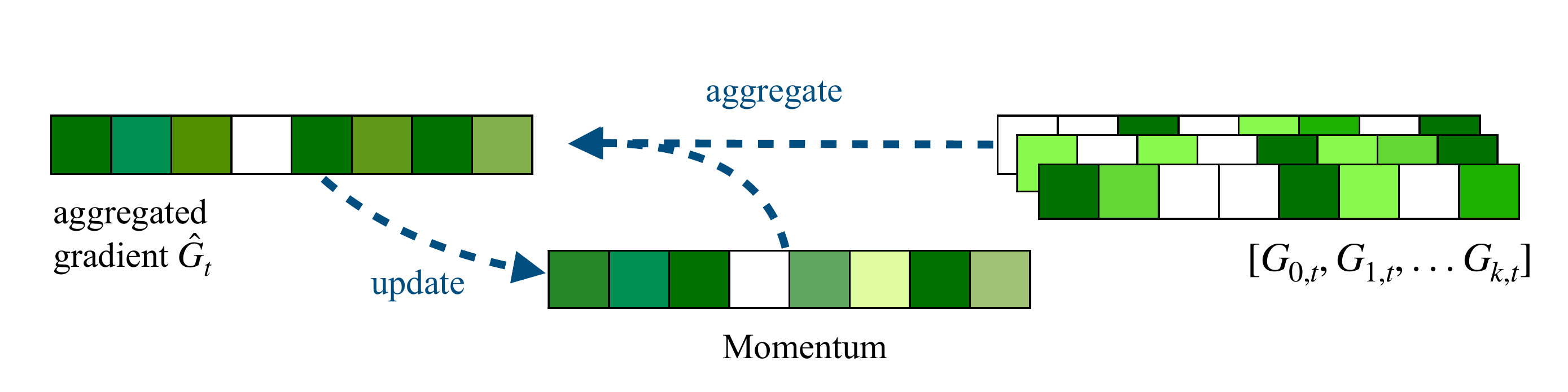}  
\vspace{-0.5em}   
\caption{Central server with global momentum.}  
\label{fig:Analysis_size}  
\vspace{-1.3em}  % 調整上下文間距
\end{figure}

Figure \ref{fig:Analysis_size} illustrate the aggergrate process with global momentum. Several works, like FetchSGD \cite{rothchild2020fetchsgd} and GSGM \cite{li2019gradient} have used global momentum on the server that gives averaged gradients. As training goes, momentum accumulate aggregate gradient each round, which making aggregated gradient nearly full size in the future rounds.

% \begin{figure}[htb]
% \centering
% % \includegraphics[width=8.5cm]{eps/server-side-agg-momentum.eps}
% \includegraphics[width=8.5cm]{pdf/server-side-momentum.pdf}
% \caption{Central aggregate server with global momentum.}
% \label{fig:Analysis_size}
% \end{figure}

In Figure \ref{fig:Analysis_size}, $T$ is total communication rounds where $t \in T$ , $k$ is client where  $k \in K$ and $\hat{G}$ is aggregated gradient.

% Hence, this issue could be mitigated if there is a more considerable similarity between $G_{k,t}$.

\subsection{Less Efficiency with Client-Side Global Momentum Compensation}\label{PF:Client-Side}
Zhoa et al. \cite{zhao2019global} took a different approach and proposed Global Momentum Compression (GMC) mechanism. GMC used global momentum to replace local momentum and achieved comparable performance. However, GMC does not consider the variance between the local gradient and the global momentum. A large variance could make the global momentum less efficient and lead to over-fitting local data, especially under low compression rates and non-IID datasets. The sizes of the gradients could have a significant disparity as the progress of each training round. Wang et al. \cite{wang2020tackling} discussed a similar issue because of the objective inconsistency of the data.

\section{Global Momentum Fusion compression scheme}
Before introducing our design of the compression process. We first dig into the major sparse compressor. The top-K sparse compression method was commented applied in many sparsification gradient compression approaches. Typically, the top-K sparse compressor creates a mask by selecting the top k\% of the most significant changing parameters of the gradient, representing the top important feature of local data on the client.

\vspace{-1em}   
\begin{equation}
    G_{k,t} \leftarrow V_{k,t \odot Mask_{topK(V_{k,t})}} \\[6pt]
\label{eq:topk}
\end{equation}

\begin{figure}[htb]
\vspace{-1em}    
\centering  
\includegraphics[width=8cm]{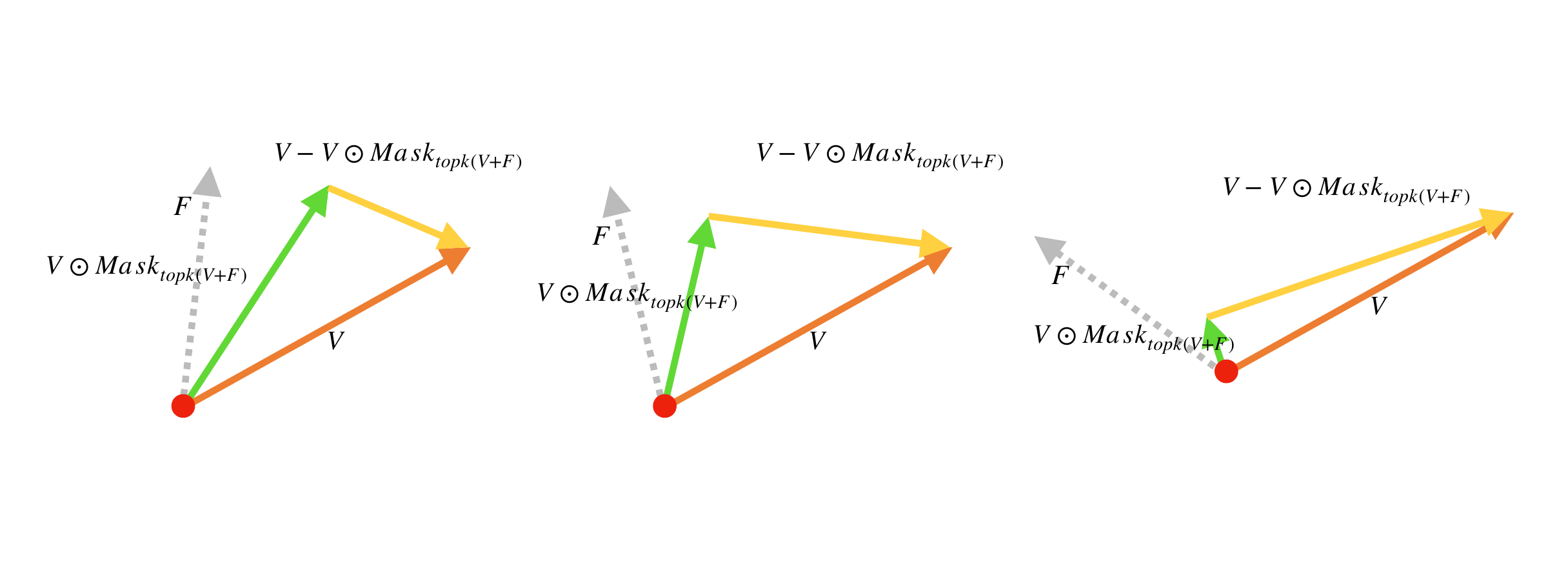}  
\vspace{-1.5em}   
\caption{Three examples of term $F$ in Top-K compressor}  
\label{fig:Top-K compressor}  
\vspace{-1.8em}  % 調整上下文間距
\end{figure}

We further design an reference of mask by adding an reference vector $F$ into top-K. In theory, these two gradients, $G^{transmit}$(green arrow) and $G^{accumulate}$(yellow arrow) in Figure \ref{fig:Top-K compressor}, are orthogonal because $G^{transmit}$ is masked by a mask, and $G^{accumulate}$ is the rest of $V$. This inspired us to design the term $F$ add with $V$ based on this property to re-weight the reference of the mask generation. Although we add a term $F$ with $V$ as shown in Figure \ref{fig:Top-K compressor}, it still stratifies the mathematical property. $F$ could be any vector that takes part in the compression selection and still satisfies the orthogonal.

We formulated our Global Momentum Fusion compression by add a fusion layer before top-K sparse compression shown in the following equation:

% \vspace{-1.3em}
% \begin{equation}
% \begin{small}
% \begin{align}
% & Z_{k,t} \leftarrow abs((1-\tau) \cdot N(V_{k,t})+\tau \cdot  N(M_{k,t}) )  \\[6pt]
% & G_{k,t} = V_{k,t} \odot Mask_{topK(Z_{k,t})} \\[6pt]
% \end{align} 
% \label{eq:GMF_compressor_policy}
% \end{small}
% \end{equation}
% % \vspace{-1em} 

\vspace{-1.5em}
\begin{equation}
\begin{small}
\centering
\begin{array}{c}
Z_{k,t} \leftarrow abs((1-\tau) \cdot N(V_{k,t})+\tau \cdot  N(M_{k,t}) )  \\[6pt]
G_{k,t} = V_{k,t} \odot Mask_{topK(Z_{k,t})} \\[6pt]
\end{array}
\label{eq:GMF_compressor_policy}
\end{small}
\end{equation}
\vspace{-1em} 

In Equation \ref{eq:topk} and \ref{eq:GMF_compressor_policy}, $V_{k,t}$ is an local compensated gradient, $G_{k,t}$ is compressed gradient, $M_{k,t}$ is the accumulated global momentum, $N$ is Normalize, $\alpha$ is the local momentum factor, $\beta$ is the global momentum factor and $\tau$ is the fusion ratio.

Additionally, inspired by FedNova \cite{wang2020tackling}, we introduced a normalized weighting method of fusion ratio to avoid domination by the local gradient and the global momentum.

\begin{figure}[htb]
\vspace{-1em}    
\centering  
\includegraphics[width=8cm]{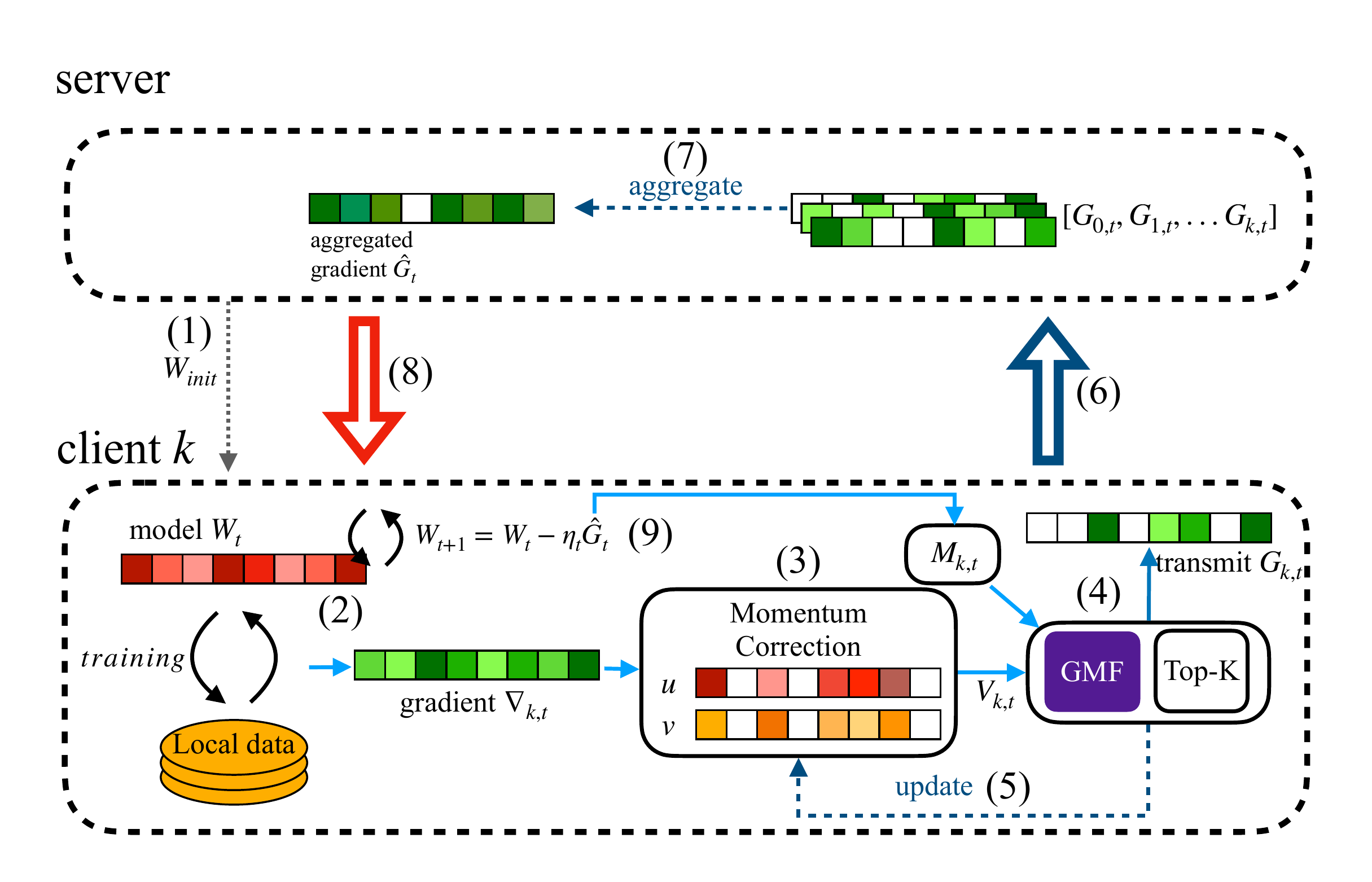}  
\vspace{-0.5em}   
\caption{DGCwGMF}  
\label{fig:DGCwGMF}  
\vspace{-1.3em}  % 調整上下文間距
\end{figure}

Next, we proposed a Deep Gradient Compression with Global Momentum Fusion (DGCwGMF) method. We adopted momentum correction from DGC as our memory compensation mechanism and implemented Global Momentum Fusion scheme in the compression policy. DGCwGMF is operates as Figure \ref{fig:DGCwGMF} and Algorithm \ref{alg:GF} illustrates the pseudocode of DGCwGMF.

\vspace{-2em} 

With the Global Momentum Fusion, we refer to the long-term momentum direction while compressing the gradient, which gives a configurable trade-off between the local gradient and the global momentum with the fusion ratio $\tau$. Before weighing them, we normalize the gradient to avoid bias caused by large variances. A smaller $\tau$ could allow the compressor selection to fit exactly to its local training data. A larger $\tau$ could waive the parameters of the gradient that differ from the global momentum. When setting $\tau > 0$, a higher similarity of the transmitted gradient between clients could have a lower expected communication overhead. When we set the fusion ratio $\tau=0$, DGCwGMF degenerates into DGC.

\vspace{-1em}   
\begin{algorithm}[H]
\begin{footnotesize}
\caption{DGCwGMF on client k}
\label{alg:GF}
\begin{algorithmic}[1]

  \State initialize $U_{k,-1} \leftarrow 0$, $V_{k,-1} \leftarrow 0$, $M_{k,-1} \leftarrow 0$
  
\State Initial base model $W_{k,0}=W_{init}$ on server and share with all clients.

\For{$t=0\dots$T}
  \State Sample batches of data $X_{k}$ from local dataset
  \State $\nabla_{k,t} = SGD(W_{k,t},X_{k})$ \{local update\}
  
  \State $U_{k,t} \leftarrow \alpha U_{k,t-1}+ \nabla_{k,t}$ \algorithmiccomment{Momentum Correction}
  \State $V_{k,t} \leftarrow V_{k,t-1} + U_{k,t}$  \algorithmiccomment{Momentum Correction}
  
\State $M_{k,t} \leftarrow \beta M_{k,t-1} + \hat{G}_{t-1}$ \algorithmiccomment{Global momentum accumulate}
\State $Z_{k,t} \leftarrow abs((1-\tau) \cdot N(V_{k,t})+\tau \cdot  N(M_{k,t}) )$ \algorithmiccomment{GMF}

\State $G_{k,t}\leftarrow  V_{k,t} \odot Mask_{topk(Z_{k,t})} $ \algorithmiccomment{Compress}

\State $U_{k,t} \leftarrow U_{k,t} \odot (1-Mask_{topk(Z_{k,t})})$  \algorithmiccomment{Memory update}
\State $V_{k,t} \leftarrow V_{k,t}\odot (1-Mask_{topk(Z_{k,t})})$  \algorithmiccomment{Memory update}
 
  \State Transmit $G_{k,t}$ to server
  \State Download aggregated $\hat{G}_{t}$ from server
  \State Update local model $W_{k,t+1} \leftarrow W_{k,t} - \eta_{t} \cdot \hat{G}_{t} $
 
\EndFor
\end{algorithmic}
\end{footnotesize}
\end{algorithm}

% \State $Z_{k,t} \leftarrow \left |(1-\tau) \cdot Normalize(V_{k,t})+\tau \cdot  Normalize(M_{k,t}) \right |$
% \input{algorithm/algorithm_gf.tex}

\section{Experiment validation}

We validate our approach with two training tasks: the image classification task in the Cifar10 dataset and the next-word prediction task in the Shakespeare dataset. To compare image classification and next-word prediction tasks, we provide a reproducible empirical experiment over two datasets containing an artificial non-IID dataset and a naturally non-IID dataset.

\subsection{Setup}

\vspace{-0.8em}

% \begin{table}[!htp] 
% \centering
% \begin{tabular}{|c|c|c}
% \hline
% & Image Classification Task & Next-Word Prediction Task    \\ \hline
% Dataset & Cifar10 & Shakespeare &  \\ \hline
% Model   & ResNet56    & \begin{tabular}[c]{@{}l@{}}RNN (single layer LSTM )\end{tabular}  \\ \hline
% \# of clients & 20                                 & 100                                                                     & \multicolumn{1}{c|}{20}        & 20            \\ \hline
% \# of rounds  & 220                                &  80                                              & \multicolumn{1}{c|}{220}       & 120           \\ \hline
% \end{tabular}
% \caption{Summary of tasks}
% \label{tab:summary_of_tasks}
% \end{table}
\begin{table}[htp]
\begin{footnotesize}
\begin{tabular}{|c|c|c|}
\hline
Task          & Image Classification      & Next-Word Prediction \\ \hline
Dataset       & Cifar10                   & Shakespeare (non-IID)              \\ \hline
Model         & ResNet56                  & RNN (single layer LSTM)  \\ \hline
\# of clients & 20                        & 100                       \\ \hline
\# of rounds  & 220                       & 80                        \\ \hline
\end{tabular}
\caption{Summary of tasks}
\label{tab:summary_of_tasks}
\end{footnotesize}
\end{table}
\vspace{-0.8em}

Table \ref{tab:summary_of_tasks} show our setting for the two task. Following the procedure in \cite{lin2018deep}, we generate 7 different Mod-Cifar10 datasets with the following Earth Moving Distance(EMD) \cite{zhao2018federated}: 0.0 (Cifar10-0), 0.48 (Cifar10-1), 0.76 (Cifar10-2), 0.87 (Cifar10-3), 0.99 (Cifar10-4), 1.18 (Cifar10-5) and 1.35 (Cifar10-6) in 20 clients to simulate realistic scenario. The EMD of the sampled clients of Shakespeare dataset is 0.1157. We set the fusion ratio $\tau$ to start from 0 and step increase to 0.6 in 10 steps. High EMD values indicate high unbalanced data. EMD=0 represents that data is balance distributed. 

\vspace{-0.6em}
% \begin{table}[!htp]
% \centering
% \scriptsize
% \begin{tabular}{lccccc}
% \toprule
% \multirow{2}{*}{Technique} &\multirow{2}{*}{MC} &\multirow{2}{*}{server-side Global Momentum} &\multicolumn{2}{c}{client-side  Global Momentum} \\\cmidrule{4-5}
% & & &Added to the compensation process & Added to the compression process\\\cmidrule{1-5}
% DGC &\checkmark &- &- &- \\\cmidrule{1-5}
% GMC &- &- &\checkmark &- \\\cmidrule{1-5}
% DGCwGM &\checkmark &\checkmark &- &- \\\cmidrule{1-5}
% DGCwGMF &\checkmark &- &- &\checkmark \\\midrule
% % \bottomrule
% \end{tabular}
% \caption{Techniques in our experiments}
% \label{tab:Technique_in_our_experiment}
% \end{table}

% Please add the following required packages to your document preamble:
% \usepackage{multirow}
\begin{table}[htp]
\begin{scriptsize}

\begin{tabular}{|c|c|c|c|}
\hline
Technique & \begin{tabular}[c]{@{}c@{}}Momentum\\ Correcton\end{tabular} & \begin{tabular}[c]{@{}c@{}}Client-­side\\ Global Momentum\end{tabular} & \begin{tabular}[c]{@{}c@{}}Server-­side\\ Global Momentum\end{tabular} \\ \hline
DGC       & v                                                            &                                                                        &                                                                        \\ \hline
GMC       &                                                              & \begin{tabular}[c]{@{}c@{}}v  (in compensation\\ process)\end{tabular} &                                                                        \\ \hline
DGCwGM    & v                                                            &                                                                        & v                                                                      \\ \hline
DGCwGMF   & v                                                            & \begin{tabular}[c]{@{}c@{}}v (in compression\\ process)\end{tabular}   &                                                                        \\ \hline
\end{tabular}
\caption{Techniques in our experiments}
\label{tab:Technique}
\end{scriptsize}
\end{table}

\vspace{-0.8em}

To analyze the performance in the term of accuracy and communication overheads with existing approaches and our work. Our experiment compares the following sparse compression methods. DGC is the method proposed by Lin Yujun et al. \cite{lin2018deep} with the momentum correction technique. We compare the current client-side global momentum work, GMC\cite{zhao2019global}. We use the same DGC algorithm to compare communication consumption and apply the server-side global momentum name as DGCwGM. Finally, DGCwGMF is the DGC method that applies the Global Momentum Fusion (GMF) scheme. Table \ref{tab:Technique}  summarizes the comparison of these techniques.

\subsection{Results of Image Classification Task}

We first compare the accuracy and communication overheads of DGCwGMF and DGCwGM in Table \ref{tab:result_of_task1}. Both DGCwGMF and DGCwGM show similar accuracy. However, DGCwGM consumes more communication overheads due to the accumulated server-side global momentum, which matches our problem formulation \ref{PF:Server-Side}.

\begin{figure}[htb]
\vspace{-1em}    
\centering  
\includegraphics[width=8cm]{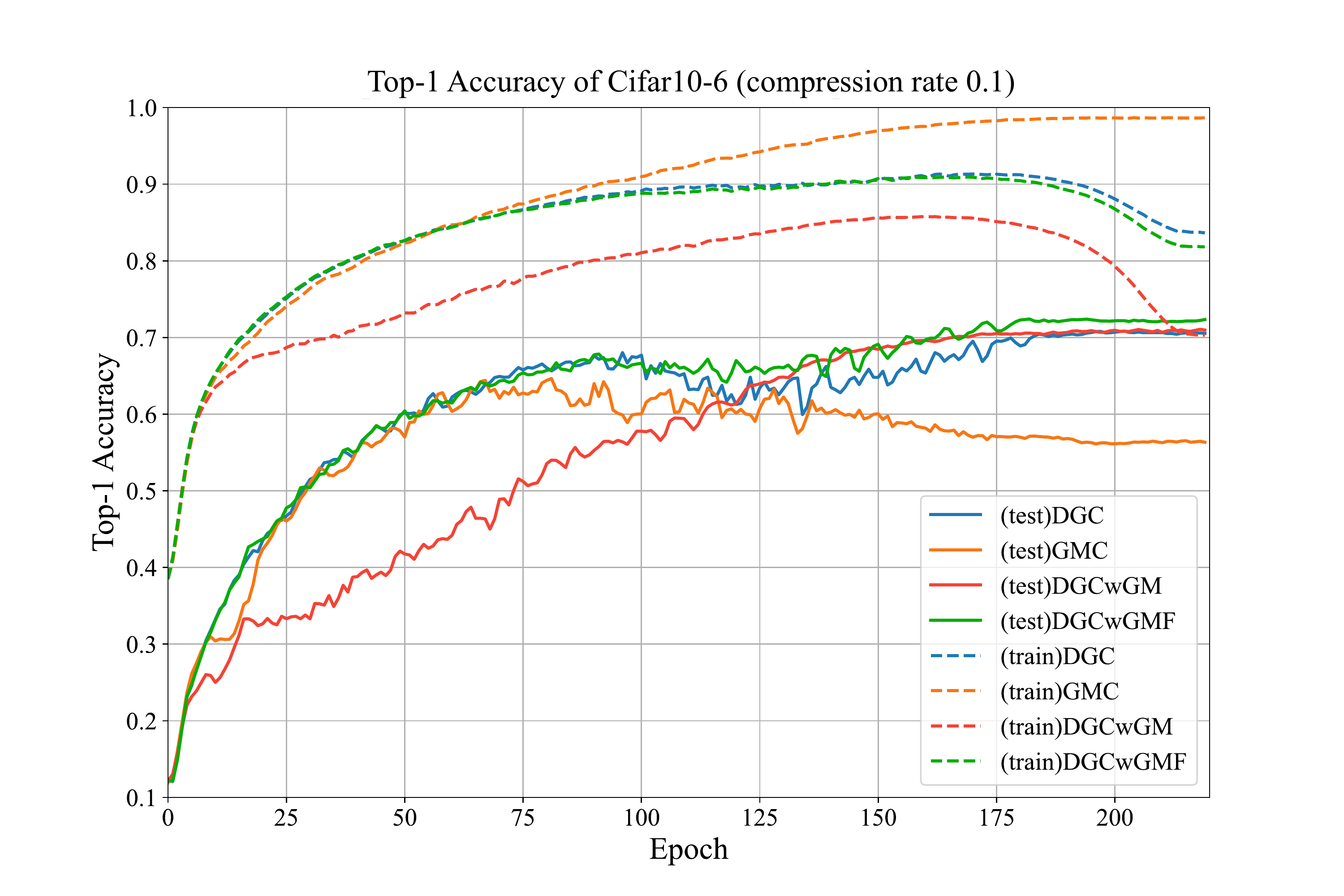}  
\vspace{-0.5em}   
\caption{Top-1 Accuracy curves on Cifar10-6}  
\label{fig:task1_top1_accuracy_6}  
\vspace{-1.3em}  % 調整上下文間距
\end{figure}

When overview our experiments, GMC reaches comparable accuracy when $EMD < 0.99$. However, in the case of compression rate=0.1 with Cifar10-6 dataset, GMC overfitted the local dataset at the end of training and degraded the performance of the model under training as shown in Figure \ref{fig:task1_top1_accuracy_6}. This issue can also match our problem formulation \ref{PF:Client-Side}.

\vspace{-0.5em}
% Please add the following required packages to your document preamble:
% \usepackage{multirow}
\begin{table}[htp]
\begin{footnotesize}
\begin{tabular}{c|ccc|cc}
\hline
Dataset                                                                          & \multicolumn{1}{c|}{Technique}      & \multicolumn{2}{c|}{Top 1 Accuracy}              & \multicolumn{2}{c}{\begin{tabular}[c]{@{}c@{}}Communication\\ Overheads\end{tabular}} \\ \hline
                                                                                 & \multicolumn{1}{c|}{DGC (Baseline)} & 0.8146 &                                         & 3.53                      &                                                           \\
                                                                                 & \multicolumn{1}{c|}{GMC}            & 0.8097 & -0.0049                                 & 3.15                      & {\color[HTML]{FE0000} -0.38}                              \\
                                                                                 & \multicolumn{1}{c|}{DGCwGM}         & 0.6238 & -0.1908                                 & 4.10                      & {\color[HTML]{000000} \textbf{+0.57}}                     \\
\multirow{-4}{*}{\begin{tabular}[c]{@{}c@{}}Cifar10-0\\ (EMD=0)\end{tabular}}    & \multicolumn{1}{c|}{DGCwGMF}        & 0.8075 & -0.0071                                 & 3.02                      & {\color[HTML]{FE0000} \textbf{-0.51}}                     \\ \hline
                                                                                 & \multicolumn{1}{c|}{DGC (Baseline)} & 0.8216 &                                         & 3.57                      &                                                           \\
                                                                                 & \multicolumn{1}{c|}{GMC}            & 0.8216 & 0                                       & 3.25                      & {\color[HTML]{FE0000} -0.32}                              \\
                                                                                 & \multicolumn{1}{c|}{DGCwGM}         & 0.7074 & -0.1142                                 & 4.12                      & \textbf{+0.55}                                            \\
\multirow{-4}{*}{\begin{tabular}[c]{@{}c@{}}Cifar10-1\\ (EMD=0.48)\end{tabular}} & \multicolumn{1}{c|}{DGCwGMF}        & 0.8095 & -0.0121                                 & 2.90                      & {\color[HTML]{FE0000} \textbf{-0.67}}                     \\ \hline
                                                                                 & \multicolumn{1}{c|}{DGC (Baseline)} & 0.8134 &                                         & 3.56                      &                                                           \\
                                                                                 & \multicolumn{1}{c|}{GMC}            & 0.7975 & -0.0159                                 & 3.29                      & {\color[HTML]{FE0000} -0.27}                              \\
                                                                                 & \multicolumn{1}{c|}{DGCwGM}         & 0.7242 & -0.0892                                 & 4.12                      & \textbf{+0.56}                                            \\
\multirow{-4}{*}{\begin{tabular}[c]{@{}c@{}}Cifar10-2\\ (EMD=0.76)\end{tabular}} & \multicolumn{1}{c|}{DGCwGMF}        & 0.8078 & -0.0056                                 & 2.83                      & {\color[HTML]{FE0000} \textbf{-0.73}}                     \\ \hline
                                                                                 & \multicolumn{1}{c|}{DGC (Baseline)} & 0.8000 &                                         & 3.56                      &                                                           \\
                                                                                 & \multicolumn{1}{c|}{GMC}            & 0.7834 & -0.0166                                 & 3.29                      & {\color[HTML]{FE0000} -0.27}                              \\
                                                                                 & DGCwGM                              & 0.7218 & -0.0782                                 & 4.12                      & \textbf{+0.56}                                            \\
\multirow{-4}{*}{\begin{tabular}[c]{@{}c@{}}Cifar10-3\\ (EMD=0.87)\end{tabular}} & \multicolumn{1}{c|}{DGCwGMF}        & 0.7970 & -0.0030                                 & 2.83                      & {\color[HTML]{FE0000} \textbf{-0.73}}                     \\ \hline
                                                                                 & \multicolumn{1}{c|}{DGC (Baseline)} & 0.7915 &                                         & 3.56                      &                                                           \\
                                                                                 & \multicolumn{1}{c|}{GMC}            & 0.7711 & -0.0204                                 & 3.29                      & {\color[HTML]{FE0000} -0.27}                              \\
                                                                                 & \multicolumn{1}{c|}{DGCwGM}         & 0.7226 & -0.0689                                 & 4.12                      & \textbf{+0.56}                                            \\
\multirow{-4}{*}{\begin{tabular}[c]{@{}c@{}}Cifar10-4\\ (EMD=0.99)\end{tabular}} & \multicolumn{1}{c|}{DGCwGMF}        & 0.7920 & {\color[HTML]{FE0000} \textbf{+0.0005}} & 2.83                      & {\color[HTML]{FE0000} \textbf{-0.73}}                     \\ \hline
                                                                                 & \multicolumn{1}{c|}{DGC (Baseline)} & 0.7712 &                                         & 3.56                      &                                                           \\
                                                                                 & \multicolumn{1}{c|}{GMC}            & 0.7146 & -0.0566                                 & 3.31                      & {\color[HTML]{FE0000} -0.25}                              \\
                                                                                 & \multicolumn{1}{c|}{DGCwGM}         & 0.7261 & -0.0451                                 & 4.12                      & \textbf{+0.56}                                            \\
\multirow{-4}{*}{\begin{tabular}[c]{@{}c@{}}Cifar10-5\\ (EMD=1.18)\end{tabular}} & \multicolumn{1}{c|}{DGCwGMF}        & 0.7706 & -0.0006                                 & 2.83                      & {\color[HTML]{FE0000} \textbf{-0.73}}                     \\ \hline
                                                                                 & \multicolumn{1}{c|}{DGC (Baseline)} & 0.7047 &                                         & 3.57                      &                                                           \\
                                                                                 & \multicolumn{1}{c|}{GMC}            & 0.5620 & -0.1427                                 & 3.34                      & {\color[HTML]{FE0000} -0.23}                              \\
                                                                                 & \multicolumn{1}{c|}{DGCwGM}         & 0.7087 & {\color[HTML]{FE0000} +0.004}           & 4.12                      & \textbf{+0.55}                                            \\
\multirow{-4}{*}{\begin{tabular}[c]{@{}c@{}}Cifar10-6\\ (EMD=1.35)\end{tabular}} & \multicolumn{1}{c|}{DGCwGMF}        & 0.7246 & {\color[HTML]{FE0000} \textbf{+0.0199}} & 2.84                      & {\color[HTML]{FE0000} \textbf{-0.73}}                     \\ \hline
\end{tabular}
\caption{Result of Image Classification Task on compression rate=0.1}
\label{tab:result_of_task1}
\end{footnotesize}
\end{table}

\vspace{-0.5em}
Next, we compare the communication overhead on Cifar10-6 for the following approaches: DGC, DGCwGMF, and DGCwGM, which have the close accuracy result, 0.7047, 0.7246, and 0.7087, respectively. When comparing the communication overhead, Table \ref{tab:result_of_task1} shows that DGCwGM costs $15.4\%$ more communication overhead than DGC. Moreover, our approach, DGCwGMF, saves $20.4\%$ of communication overhead compared with DGC and reaches higher accuracy, which proves GMF could actually save more communication overhead and less accuracy loss or improve accuracy.

\begin{figure}[htb]
\vspace{-1em}    
\centering  
\includegraphics[width=8cm]{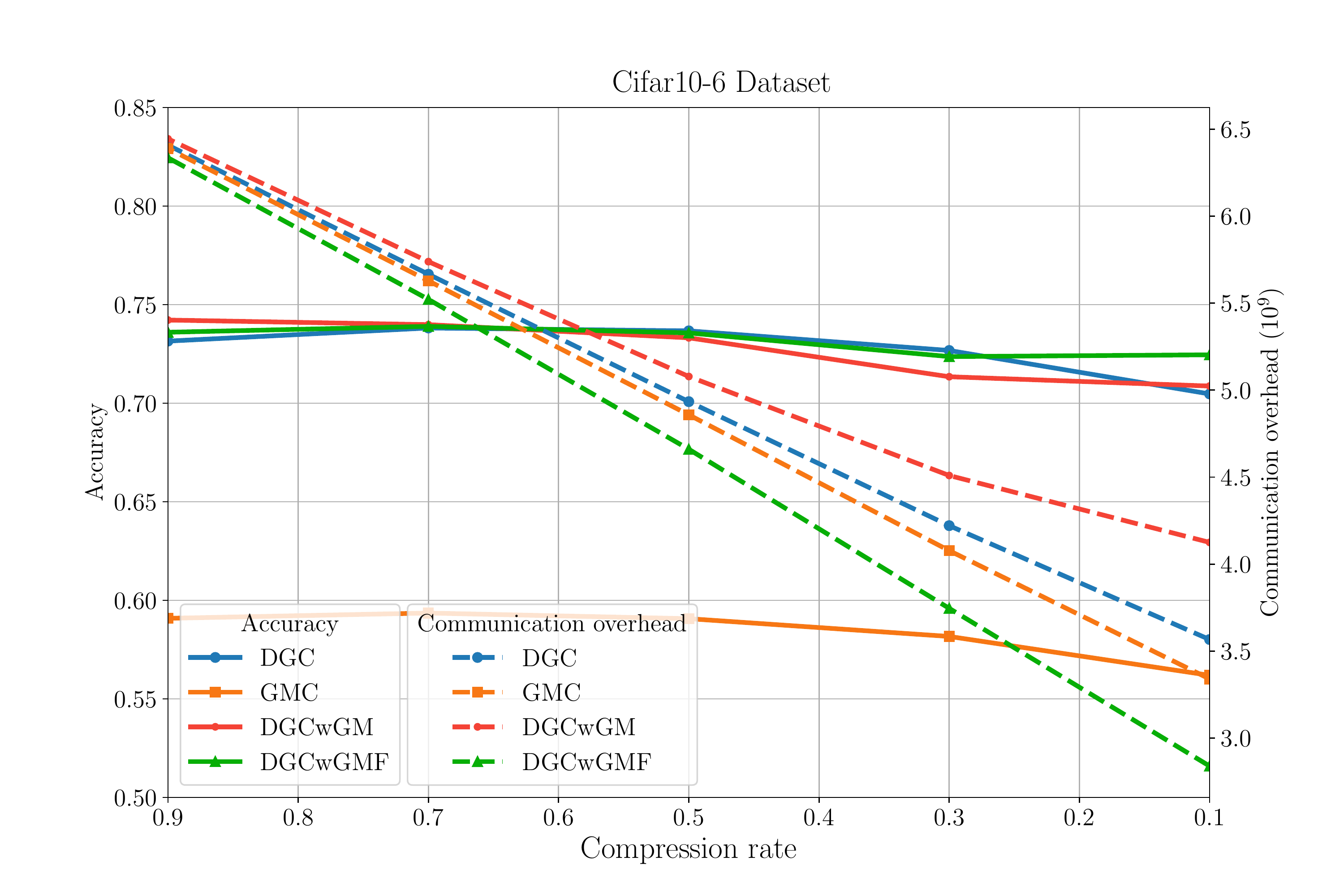}  
\vspace{-0.5em}   
\caption{Comparison of Test Accuracy and communication overhead over compression rate on Cifar10-6}  
\label{fig:task1_comparison_accuracy_compression_rate}  
\vspace{-1.3em}  % 調整上下文間距
\end{figure}

Finally, we provide experiments of accuracy and communication overhead over different compression rate from 0.1 to 0.9 on Cifar10-6. In Figure \ref{fig:task1_comparison_accuracy_compression_rate}, the experiments result show that all techniques lose performance with the compression rate decrease. However, DGCwGMF maintains less performance loss under a low compression rate. DGCwGMF also decreases faster than other approaches on the communication overhead curve, making it ideal for further compression.

\subsection{Results of Next-word Prediction Task}

In Next-Word Prediction Task, Table \ref{tab:result_of_task2} shows that the DGCwGM has more communication overhead consumption than DGC, which also proves our problem formulation \ref{PF:Server-Side}, that momentum at the server-side requires more communication overhead consumption. Next, the table also show that GMC fails to converge. However, the other work, DGC, DGCwGMF, and DGCwGM, have very close accuracy.

\vspace{-0.8em}
% Please add the following required packages to your document preamble:
% \usepackage{multirow}
\begin{table}[htp]
\begin{footnotesize}

\begin{tabular}{c|c|cc|cc}
\hline
Dataset                                                                              & Technique      & \multicolumn{2}{c|}{Top 1 Accuracy}   & \multicolumn{2}{c}{\begin{tabular}[c]{@{}c@{}}Communication\\ Overhead\end{tabular}} \\ \hline
                                                                                     & DGC (Baseline) & 0.419 &                               & 2.72                     &                                                           \\
                                                                                     & GMC            & 0.347 & -0.072                        & 1.95                     & {\color[HTML]{FE0000} -0.77}                              \\
                                                                                     & DGCwGM         & 0.421 & {\color[HTML]{FE0000} +0.002} & 3.02                     & \textbf{+0.3}                                             \\
\multirow{-4}{*}{\begin{tabular}[c]{@{}c@{}}Shakespeare\\ (EMD=0.1157)\end{tabular}} & DGCwGMF        & 0.419 & \textbf{0}                    & 2.38                     & {\color[HTML]{FE0000} \textbf{-0.34}}                     \\ \hline
\end{tabular}

\caption{Result of Next-Word Prediction Task on compression rate=0.1}
\label{tab:result_of_task2}
\end{footnotesize}
\end{table}
\vspace{-0.8em}

Besides GMC, the result of DGC, DGCwGMF, and DGCwGM under compression rate=0.1 also shows that DGCwGMF consumes less communication overhead and reaches only $0.002$ accuracy loss compared with DGCwGM. Table \ref{tab:result_of_task2} shows that DGCwGM costs $11\%$ more communication overhead than DGC, and DGCwGMF saves $12.5\%$ more communication overhead than DGC.

\begin{figure}[htb]
\vspace{-1em}    
\centering  
\includegraphics[width=8cm]{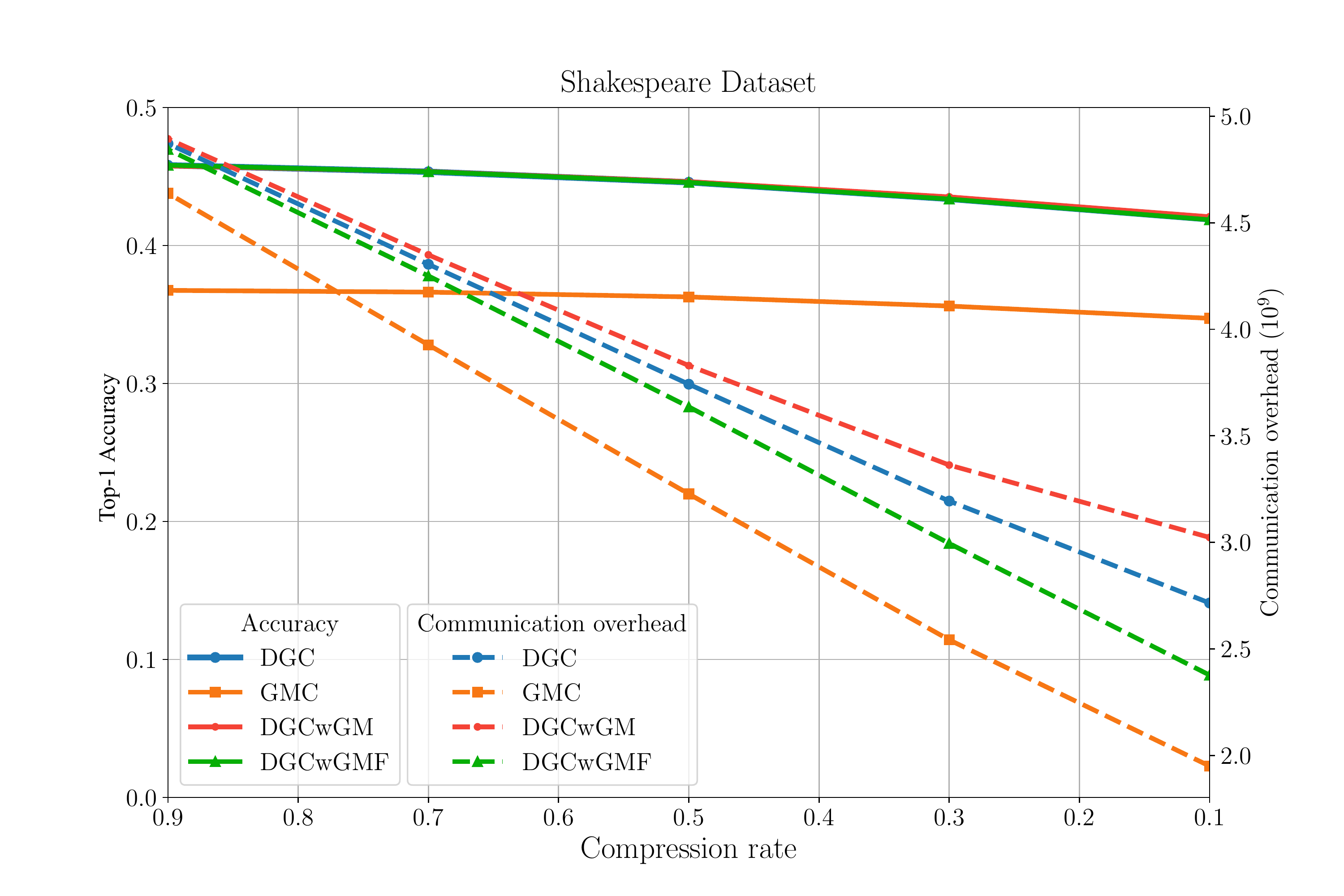}  
\vspace{-0.5em}   
\caption{Comparison of Test Accuracy and communication overhead over compression rate on Shakespeare}  
\label{fig:task2_comparison_accuracy_compression_rate}  
\vspace{-1.3em}  % 調整上下文間距
\end{figure}

Finally, we analyze the performance between different compression rates. Figure \ref{fig:task2_comparison_accuracy_compression_rate} shows that the performance is nearly the same with other works but requires fewer communication overheads.

\section{Conclusion}

We proposed a gradient compression with a compensation scheme called Deep Gradient Compression with Global Momentum Fusion (DCGwGMF) to reduce FL communication overhead while maintaining comparable model performance despite non-IID data. We designed and conducted comparative experiments to prove the proposed DCGwGMF provides the best performance with low compression rates in high EMD datasets. Specifically, our experiments showed that DGCwGMF has less accuracy loss in the non-IID dataset and $20.4\%$ fewer communication overheads than DGC in the image classification task. DGCwGMF reduces $12.5\%$ of communication overheads and maintains the same accuracy as vanilla DGC for speech recognition tasks.

\vfill\pagebreak

% \section{REFERENCES}
% \label{sec:refs}

% List and number all bibliographical references at the end of the
% paper. The references can be numbered in alphabetic order or in
% order of appearance in the document. When referring to them in
% the text, type the corresponding reference number in square
% brackets as shown at the end of this sentence \cite{lin2018deep}. An
% additional final page (the fifth page, in most cases) is
% allowed, but must contain only references to the prior
% literature.

% References should be produced using the bibtex program from suitable
% BiBTeX files (here: strings, refs, manuals). The IEEEbib.bst bibliography
% style file from IEEE produces unsorted bibliography list.
% -------------------------------------------------------------------------
\bibliographystyle{IEEEbib}
% \bibliography{strings,refs}
\bibliography{reference}

\begin{thebibliography}{1}

\bibitem{mcmahan2017communication}
Brendan McMahan, Eider Moore, Daniel Ramage, Seth Hampson, and Blaise~Aguera
  y~Arcas,
\newblock ``Communication-efficient learning of deep networks from
  decentralized data,''
\newblock in {\em Artificial intelligence and statistics}. PMLR, 2017, pp.
  1273--1282.

\bibitem{xu2020compressed}
Hang Xu, Chen-Yu Ho, Ahmed~M Abdelmoniem, Aritra Dutta, El~Houcine Bergou,
  Konstantinos Karatsenidis, Marco Canini, and Panos Kalnis,
\newblock ``Compressed communication for distributed deep learning: Survey and
  quantitative evaluation,''
\newblock Tech. {R}ep., 2020.

\bibitem{lin2018deep}
Yujun Lin, Song Han, Huizi Mao, Yu~Wang, and Bill Dally,
\newblock ``Deep gradient compression: Reducing the communication bandwidth for
  distributed training,''
\newblock in {\em International Conference on Learning Representations}, 2018.

\bibitem{aji2017sparse}
Alham Aji and Kenneth Heafield,
\newblock ``Sparse communication for distributed gradient descent,''
\newblock in {\em EMNLP 2017: Conference on Empirical Methods in Natural
  Language Processing}. Association for Computational Linguistics (ACL), 2017,
  pp. 440--445.

\bibitem{zhao2019global}
Shen-Yi Zhao, Yin-Peng Xie, Hao Gao, and Wu-Jun Li,
\newblock ``Global momentum compression for sparse communication in distributed
  sgd,''
\newblock {\em arXiv preprint arXiv:1905.12948}, 2019.

\bibitem{rothchild2020fetchsgd}
Daniel Rothchild, Ashwinee Panda, Enayat Ullah, Nikita Ivkin, Ion Stoica,
  Vladimir Braverman, Joseph Gonzalez, and Raman Arora,
\newblock ``Fetchsgd: Communication-efficient federated learning with
  sketching,''
\newblock in {\em International Conference on Machine Learning}. PMLR, 2020,
  pp. 8253--8265.

\bibitem{li2019gradient}
Chengjie Li, Ruixuan Li, Haozhao Wang, Yuhua Li, Pan Zhou, Song Guo, and Keqin
  Li,
\newblock ``Gradient scheduling with global momentum for non-iid data
  distributed asynchronous training,''
\newblock {\em arXiv preprint arXiv:1902.07848}, 2019.

\bibitem{wang2020tackling}
Jianyu Wang, Qinghua Liu, Hao Liang, Gauri Joshi, and H~Vincent Poor,
\newblock ``Tackling the objective inconsistency problem in heterogeneous
  federated optimization,''
\newblock {\em Advances in neural information processing systems}, vol. 33, pp.
  7611--7623, 2020.

\bibitem{zhao2018federated}
Yue Zhao, Meng Li, Liangzhen Lai, Naveen Suda, Damon Civin, and Vikas Chandra,
\newblock ``Federated learning with non-iid data,''
\newblock {\em arXiv preprint arXiv:1806.00582}, 2018.

\end{thebibliography}

\end{document}